\documentclass[review]{elsarticle}
\usepackage[top=3cm,bottom=3cm,left=2.2cm,right=2.2cm]{geometry}
\usepackage{lipsum}
\makeatletter
\usepackage{lineno,hyperref}
\usepackage{textcomp}
\usepackage{amsfonts}
\usepackage{amsmath}
\usepackage{amssymb}
\usepackage{rotfloat}
\biboptions{numbers,sort&compress}
\usepackage{graphicx}
\usepackage{dcolumn}
\usepackage{bm}
\usepackage{array}
\usepackage{setspace}
\usepackage{xcolor}

\def\ps@pprintTitle{%
	\let\@oddhead\@empty
	\let\@evenhead\@empty
	\def\@oddfoot{}%
	\let\@evenfoot\@oddfoot}
\makeatother

\begin{document}

\begin{frontmatter}

\title{Origin of the Curie-von Schweidler law and the fractional capacitor
	from time-varying capacitance}

\author{Vikash Pandey}
\address{School of Interwoven Arts and Sciences, Krea University, Sri City, India}

\ead{vikash.pandey@krea.edu.in}

\begin{abstract}
Most dielectrics of practical purpose exhibit memory and are described
by the century-old Curie--von Schweidler law. Interestingly, the Curie--von
Schweidler law is the motivation behind an unconventional circuit
component called \textit{fractional capacitor} which due to its power-law
property is extensively used in the modeling of complex dielectric media. Unfortunately, the empirical nature of the Curie--von Schweidler law plagues
the applications of the fractional capacitor. Here, we derive the Curie--von Schweidler
law from a series combination of a resistor and a capacitor with a linear time-varying
capacitance. This may possibly be its \textit{first} derivation from physical
principles. However, this required a modification of the classical charge--voltage relation of a capacitor to account for the time-varying capacitance. The limitation of the classical charge--voltage relation and its subsequent modification are justified using appropriate circuit modeling. Consequently, the parameters of the Curie--von Schweidler
law and the fractional capacitor gain physical interpretation. The Debye response of dielectrics emerges naturally from the limiting case of the power-law response at short timescales. The
obtained results are validated by matching them with the published experimental reports.
\end{abstract}


\end{frontmatter}

\vspace{20pt}

\textbf{NOTE:} \texttt{The first version of this manuscript was uploaded
as \href{https://arxiv.org/abs/2006.06073}{arXiv:2006.06073} on June
10, 2020. }
\texttt{The peer-reviewed version of the manuscript is published in
the Journal of Power Sources, Vol.~532, Pages 231309, 2022.} 
DOI: 
\href{https://doi.org/10.1016/j.jpowsour.2022.231309}{https://doi.org/10.1016/j.jpowsour.2022.231309}. 

\texttt{The published version of the manuscript is available online
at 
\\\href{https://www.sciencedirect.com/science/article/pii/S0378775322003238}{https://www.sciencedirect.com/science/article/pii/S0378775322003238}.}

\textcolor{blue}{\textbf{This document is an e-print which may differ
in, e.g. pagination, referencing styles, figure sizes, and typographic
details.}}

\newpage

\section{Introduction\label{sec:introd}}

Most electrical appliances of daily use contain capacitors to store
electrical energy. The storage of energy occurs through the charge
accumulation on the capacitor plates. This is facilitated by the dielectric
media in the capacitor that exhibits high polarizability under the
action of an applied electrical field. However, a lag in the polarization
leads to dielectric relaxation. Since an insight into the optical
and electrical properties of a dielectric is possible from its relaxation
behavior, the study of dielectric relaxation has attracted interest from material
scientists and electrical engineers alike for more than a century
\cite{Jonscher1983}. 

According to the classical theory of Debye relaxation, the current
response, $I_{D}\left(t\right)$, of an \textit{ideal} dielectric
to an input step voltage is a memoryless exponential decay, i.e.,~$I_{D}\left(t\right)\propto\exp\left(-t/\tau_{D}\right)$, where,
$t$ is the time, and $\tau_{D}$ is the Debye relaxation time constant.
The response constitutes a sum of responses from a non-interacting population
of dipoles, but that could only be justified for relatively dilute dipolar
systems in which the inter-particle distances are large. In contrast,
most dielectrics are solids and so mutual interactions cannot be ignored. 
Solid dielectrics exhibit memory, i.e.,~they remember their past excitations
because of non-Debye relaxation mechanisms. Such a memory-laden behavior is often described
using power-laws~\cite{Jonscher1977a} which was first experimentally
inferred by Curie in 1889~\cite{Curie1889} and later rediscovered
by von Schweidler in 1907~\cite{Schweidler1907}. The Curie--von Schweidler
(CvS) law describes the power-law decay of a depolarizing current
in a dielectric that is subjected to a step DC voltage as:
\begin{equation}
 I\left(t\right)=a\left(\frac{t}{\tau}\right)^{-\alpha},\label{eq:cvslaw}
\end{equation}
where $a$ has the dimension of current, $\tau$ is the characteristic
relaxation time constant, and $0<\alpha<1$ is the decay constant. Some other equations have also been
proposed to describe dielectric relaxation, such as Cole--Cole~\cite{Cole1941}, Cole--Davidson
\cite{Khamzin2013a}, Havriliak--Negami~\cite{Schueoller1994}, and
Kohlrausch--Williams--Watts~\cite{Kriza1986,Bunde1997}, but in comparison
to Eq.~\ref{eq:cvslaw}, other empirical equations are
relatively difficult to curve-fit~\cite{Grosse2014}. Consequently, the
CvS law arguably remains the preferred one among all. Also, the CvS law is
weakly influenced by the physical structure, geometric configurations,
and chemical bonding of the material. Besides,
it is largely independent of the nature of the polarizing species,
be they hopping electrons, ions, or dipoles~\cite{Jonscher1977a}.  The law holds regardless of
changes in the temperature except when the change is radical. Since the CvS
law is sufficiently good and surprisingly better in most cases, it was declared as the universal dielectric response (UDR)~\cite{Jonscher1983,Jonscher1977a}. 
Further, the Fourier transform property of the power-law, ${\mathcal{F}}\left[t^{-\alpha}\right]=\Gamma\left(1-\alpha\right)\left(i\omega\right)^{\alpha-1}, \mbox{for }t>0$,
is used to alternatively express the CvS law as, $\chi\left(\omega\right)\propto\omega^{\alpha-1}$, where $\chi$ is the complex electric susceptibility, $\omega$ is the angular frequency, $i$ is the imaginary number, and $\Gamma\left(\cdot\right)$
is the Euler Gamma function~\cite{Jonscher1977a}. The validity of
the relation is established in as many as eight
to ten decades of frequency~\cite{Jonscher1983,Jonscher2001,Lunkenheimer2003}. It is worthwhile to mention that the CvS law is plagued by a mathematical singularity at extremely short timescales which also mirrors at very
high $\left(>\text{THz}\right)$ frequencies. The underlying physical reasons behind this are the quantum effects due to phonon and lattice
vibrations which dominate at short timescales~\cite{Jonscher1983}. In the remaining part of this manuscript we focus
on the CvS law expressed by Eq.~\ref{eq:cvslaw} for two reasons. First, because of its historical context, and second, because it is relatively
easier to measure current than susceptibility in experiments.

Probably, the first attempt to uncover the physics underlying the CvS
law was motivated from a pure mathematical result that equals power
law with an infinite weighted sum of Debye relaxation responses~\cite{Schweidler1907}.
Since the respective circuit model consists of an infinite ladder
network of resistors and capacitors~\cite{Gross1990}, it leads to
an infinite system of first order differential equations such that each of those differential
equations yields a weighted exponential  with a distinct relaxation
time constant as its solution~\cite{Jonscher1983}. However, irrespective
of the mathematical consistency, the breadth and the form of any such
relaxation distribution are difficult to comprehend~\cite{Guo1983,Schaefer1996}.
Since they neither aid nor advance the understanding, the reasoning
is arbitrary and may even be considered superfluous~\cite{Jonscher1977a,Cole1941}.
Even though von Schweidler had followed the same approach~\cite{Schweidler1907},
it is seen as a mere attempt to reconcile the observed memory exhibits
with the desired yet incompatible memoryless Debye relaxation processes.
This is understandable  because back then it was almost counter-intuitive
to acknowledge that matter could have a memory. 

An alternative mechanism is many body interactions that manifests
from the nonlocal reciprocal interactions between the polarizing species
and the matter lattice in which they move~\cite{Jonscher1977a,Jonscher2001}, but
it is impossible to study them at atomic and molecular levels because
of the resulting mathematical complexities~\cite{Jonscher1983,Uchaikin2009}. Also,
the values of the density and the magnitude of the polarizing species
are rarely available~\cite{Despotuli2014}. Some more attempts inspired
by the common observation of self-similarity in power-laws and fractal
geometry have been made~\cite{Leyderman2000,Novikov2001,Raicu2001,Stanislavsky2007,Khamzin2013b}.
Unfortunately, those attempts and a few others~\cite{Ngai1979a,Bagchi1990}
are all specifically tailored to satisfy the observed power-law behavior
without giving any insight into the mechanism that governs the UDR.
This is evident from the lack of a physical interpretation of the
parameters, $a$, $\tau$, and $\alpha$. Unfortunately, the
status of the UDR has reduced to an empirical relation that can only be used to curve-fit
the experimental data~\cite{Kotecki1999,Cang2003,Miranda2009,Popov2012a,Ma2019}. 

The undeniable fact that the CvS law is ubiquitous indicates that
a more fundamental yet a universal mechanism is at play. In fact,
its theoretical description is considered as one of most \textit{fundamental}
problems in physics~\cite{Jonscher2001,Despotuli2014,Maass1991,Tarasov2008,Stanislavsky2017}.
Therefore, the quest for an encompassing interpretation becomes imperative and serves as a motivation for this manuscript. Coincidentally, the power-law anchored in Eq.~\ref{eq:cvslaw} is inherent in
the definition of a fractional derivative. The Caputo definition for
the fractional derivative of a causal, continuous function, $f\left(t\right)$,
is the convolution of an integer-order derivative with a power-law
memory kernel, $\phi_{\alpha}\left(t\right)$, as~\cite{Mainardi2010}:
\begin{equation}
 \frac{d^{\alpha}}{dt^{\alpha}}f\left(t\right)\overset{\text{def.}}{=}\dot{f}\left(t\right)\ast\phi_{\alpha}\left(t\right),\text{ }\phi_{\alpha}\left(t\right)=\frac{t^{-\alpha}}{\Gamma\left(1-\alpha\right)},\text{ }0<\alpha<1,\label{eq:frac_derivat}
\end{equation}
where the number of over-dots represent the order of differentiation
with respect to time, $t$. Although fractional derivatives are expressed in the integro-differentiable form,
their Fourier transform property, $\mathcal{F}\left[d^{\alpha}f\left(t\right)/dt^{\alpha}\right]=\left(i\omega\right)^{\alpha}\boldsymbol{f}\left(\omega\right)$, affirm that they are a mere generalization of the regular integer-order derivatives. For a negative value of $\alpha$, Eq.~\ref{eq:frac_derivat}  corresponds to a fractional integral.

The motivation to investigate the UDR using fractional derivatives
is three-fold. \textit{First}, the kernel, $\phi_{\alpha}\left(t\right)$,
from Eq.~\ref{eq:frac_derivat}, is identified as the power
law that characterizes the CvS law described by Eq.~\ref{eq:cvslaw}. In addition, since $\mathcal{F}\left[\phi_{\alpha}\left(t\right)\right]$
yields the UDR in the respective frequency domain, fractional derivatives appear as a natural tool for their investigation~\cite{Garrappa2016,Tarasov2008}. Surprisingly, the Mittag-Leffler functions which constitute the
basic solution of a fractional differential equation have been used in the modeling of anomalous dielectric relaxations~\cite{Garrappa2016}. Further, it is known that the evolution of macroscopic memory from the underlying
microscopic
mechanisms is very complex and so fractional derivatives
are almost inevitable~\cite{Stanislavsky2000}. It should be noted that fractional
derivatives inherently include a multiscale generalization that accounts
for the nonlocal interactions~\cite{Magin2010,Pandey2016}. Interestingly, Westerlund had even
claimed that Nature works with fractional derivatives~\cite{Westerlund1991}. 

\textit{Second}, electrical engineers have exploited the universality
of the UDR in the modeling of biological membranes~\cite{Raicu1999,Ionescu2009,Moreles2017}, differentiation of cancerous tissues from healthier tissues~\cite{Laufer2010}, and designing of new energy storage devices,
example supercapacitors~\cite{Kotecki1999,Allagui2018b,Band2019}.
The respective circuit models include an unconventional component
called fractional capacitor whose current--voltage relation is given as:
\begin{equation}
 I_{C_{f}}\left(t\right)=C_{f}\frac{d^{\alpha}V_{C_{f}}\left(t\right)}{dt^{\alpha}},\text{  }  C_{f}=G\tau^{\alpha}, \label{eq:frac_cap}
\end{equation}
where $I_{C_{f}}$ and $V_{C_{f}}$ are the current
and the voltage respectively, $C_{f}$ is the pseudocapacitance, $G$ is the electrical conductance, $t$ is the time, and $\tau$ and $\alpha$ are the same as in Eq.~\ref{eq:cvslaw}. Since a fractional
capacitor exhibits an interplay between a resistance and a capacitor for values of $\alpha$ lying between $0$ and $1$, it may be seen as an electrical analogue of a fractional dashpot from non-Newtonian rheology
\cite{Garrappa2016,Pandey2016a}. In light of Eq.~\ref{eq:frac_derivat},
the memory kernel anchored in Eq.~\ref{eq:frac_cap}, is inferred to be the CvS law. The fractional
capacitor has a characteristic frequency-independent fixed phase angle,
$\left|\alpha\pi/2\right|$, which means the energy lost per cycle
is the same fraction of the energy stored at all frequencies. This
is the same as the constant ratio of, $\Im\left[\chi\left(\omega\right)\right]/\Re\left[\chi\left(\omega\right)\right]$, where $\Re$ and $\Im$ denote the real and imaginary parts of a complex number. For these reasons, the fractional capacitor is also called the \textit{constant phase element}. Although Jonscher had already claimed the CvS law to be the response
of a ``universal'' capacitor~\cite[p.~87]{Jonscher1983},
Westerlund formally introduced it as a fractional capacitor
to the electrical engineers community~\cite{Westerlund1991,Westerlund1994}.
The first affirmation regarding the universality of the CvS law can be traced back to the Cole-brothers who were probably motivated from Gemant's work on rheology~\cite{Gemant1936} whom they even cited as Ref.~{[42]}
in their work~\cite{Cole1941}.
The versatility of a fractional capacitor ensures requirement of fewer parameters in the modeling of complex electrical behavior. Although their use provides a better curve-fit than those from integer-order derivatives,  they suffer from an inherent drawback that stems from the lack of a physical interpretation of $\alpha$
\cite{Podlubny2002,Machado2015}.

\textit{Third}, power-law behaviors are common in acoustics too, and
their investigation using fractional derivatives has led to the development
of the field of fractional viscoelasticity~\cite{Mainardi2010,Bonfanti2020}. Surprisingly, similar to the case of dielectric modeling, the application of fractional derivatives in
rheological modeling usually had their motivation from the continuum
of multiple relaxation processes~\cite{Nasholm2011}. In contrast, few researchers have recently derived fractional differential equations from physical
principles that have also led to the physical interpretation of the fractional
order~\cite{Pandey2016a,Pandey2016b,Pandey2016c,Holm2016,Sun2018,Holm2019}. Some of those findings~\cite{Pandey2016a} have been independently verified in experiments as well~\cite{Hofer2019,Hofer2020,Yang2020}.

The rest of the article is organized as follows.  \ref{sec:Modification_classical_relation} is divided into two subsections. In its first subsection, we first prove the limitation of the classical charge--voltage relation for a time-varying capacitor. In particular, it is shown that the classical relation, $Q\left(t\right)=CV\left(t\right)$, that relates the charge, $Q$, with the capacitance, $C$, and the voltage, $V$, is not applicable for capacitors with a time-varying capacitance, $C\left(t\right)$. The expression for the current, $dQ/dt$, that is subsequently obtained following the substitution of $C$ by  $C\left(t\right)$ in the classical relation corresponds to a non-equivalent circuit. The equivalence of the circuit is restored in the second subsection of  \ref{sec:Modification_classical_relation} where a modified charge--voltage relation is proposed, and also suitably justified. The modified charge--voltage relation is then used to derive the universal dielectric response, the CvS law, in  \ref{sec:Deriv_CvS_law}. Lastly, the implications of this work are discussed in  \ref{sec:Discussion}.

\section{Revision of classical charge-voltage relation\label{sec:Modification_classical_relation}} \label{sec:Rev}

An efficient way to represent a capacitor's memory is to assume a time-varying capacitance, $C\left(t\right)$.  Such an assumption has been used in the study of solid state devices~\cite{Lee1993,Brauer1998}, time-varying
storage components~\cite{Biolek2007,Richards2012}, energy accumulation~\cite{Mirmoosa2019}, brain microvasculature~\cite{Jo2015}, and biomimetic membranes~\cite{Najem2019}. Most of these cited references use the following
charge--voltage equation,
\begin{equation}
Q\left(t\right)=C\left(t\right)V\left(t\right),\label{eq:charge_relation}
\end{equation}
as a starting point in their studies. The equation is directly motivated from the classical charge--voltage relation of a capacitor, $Q=CV$, where, $Q$ is the accumulated charge, $C$ is the constant capacitance, and $V$ is the applied voltage. The current is then obtained from Eq.~\ref{eq:charge_relation} following the product rule of differentiation:
\begin{equation}
I\left(t\right)=\dot{Q}\left(t\right)=C\left(t\right)\dot{V}\left(t\right)+V\left(t\right)\dot{C}\left(t\right).\label{eq:current_relation}
\end{equation}

\subsection{Limitation of classical charge-voltage relation}\label{sec:Reva}

\begin{figure}[h!]
\begin{centering}
\includegraphics[width=1\columnwidth]{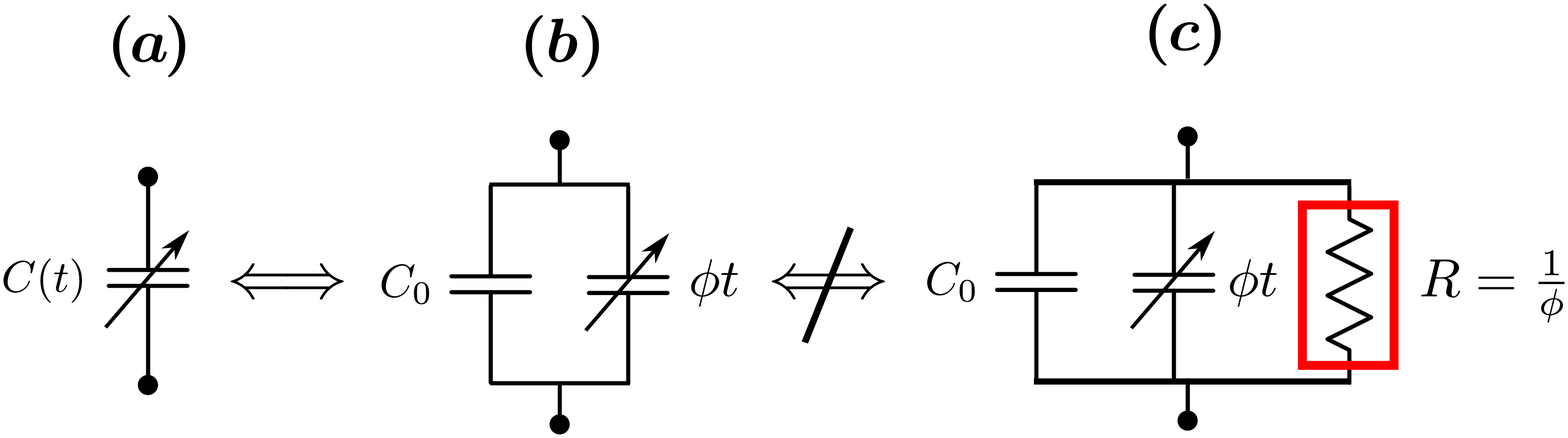}
\par\end{centering}
\caption{(a) Conventional symbol for a time-varying capacitor with a time-varying capacitance, $C\left(t\right)$. (b) The equivalent model of (a), assuming a linearly time-varying capacitance, $C\left(t\right)=C_{0}+\phi t$. (c) The non-equivalent model with an undesired resistor that emerges from the application of the classical relation,  $Q\left(t\right)=C\left(t\right)V\left(t\right)$, to (b).}
\begin{centering}
\label{Time_capacitor}
\par\end{centering}
\end{figure}

The time-varying capacitance of a capacitor can be expressed as,
\begin{equation}
C\left(t\right)=C_{0}+C_{\phi}\left(t\right),\label{eq:time_var_cap}
\end{equation} 
where  $C_{0}$ is the constant geometric capacitance, and $C_{\phi}\left(t\right)$ is the time-varying part of the capacitance due to the dielectric media present in the capacitor, see Eq.~(2) in
Ref.~\cite{Westerlund1991}. Further, we assume the time-varying part of the capacitance to be a linear function in $t$, given by, $C_{\phi}\left(t\right)=\phi t$, where the constant, $\phi>0$. Since the capacitances of the individual capacitors add in parallel
circuits, the equivalent circuits are shown in Figs.~\ref{Time_capacitor} (a) and (b).
On substituting Eq.~\ref{eq:time_var_cap} in  Eqs.~$\left(\ref{eq:charge_relation}\right)$ and $\left(\ref{eq:current_relation}\right)$, we have the following:
\begin{equation}
Q\left(t\right)=\left(C_{0}+\phi t\right)V\left(t\right)\text{, and }\label{eq:curie_charge}
\end{equation}

\begin{equation}
I\left(t\right)=C_{0}\dot{V}+\phi t\dot{V}+V\phi.\label{eq:curie_current}
\end{equation}
On carefully observing the three additive terms on the right hand side of Eq.~\ref{eq:curie_current}, we find that the first
two terms correspond to the currents that flow through the capacitors of capacitances, $C_{0}$ and $C_{\phi}$, respectively, but the third term is the Ohmic current that flows through a resistor of an equivalent
resistance, $1/\phi$. Since currents add in the parallel branches, Eq.~\ref{eq:curie_current}, effectively corresponds to a parallel combination of the three elements as shown in  \ref{Time_capacitor}(c). Clearly, this is not
equivalent to~\ref{Time_capacitor}(b). This anomaly may also be verified as follows. On imposing the initial condition at time, $t=0$, we have from Eq.~\ref{eq:curie_current}, the current, $I_{0}=C_{0}\dot{V}+V\phi$,
instead of the expected, $I_{0}=C_{0}\dot{V}$. Further, if $V$ is a constant, i.e.,~$\dot{V}=0$, then the current flowing through the capacitor should be zero, but on the contrary, Eq.~\ref{eq:curie_current} predicts an undesired ohmic current,
$V\phi$. The
underlying reason behind the non-equivalence of  Figs.~\ref{Time_capacitor} (b) and  \ref{Time_capacitor} (c) is that the traditional charge--voltage relation, Eq.~\ref{eq:charge_relation}, assumes a linear time-invariant system, i.e.,~$Q\left(t\right)=f\left(V\left(t\right)\right)$. In contrast, a time-varying capacitor is a time-varying system, $Q\left(t\right)=f\left(V\left(t\right), t\right)$, i.e.,~the capacitor remembers the applied voltage that it was subjected to, in the past~\cite{Westerlund1994}. The classical
relation  leads to a term-by-term multiplication of $C\left(t\right)$ and $V\left(t\right)$, and therefore it does not take the capacitor memory into account.

Here it is necessary to emphasize that even though the classical relation is only applicable for capacitors with a constant capacitance~\cite{Westerlund1994}, Eqs.~$\left(\ref{eq:charge_relation}\right)$ and $\left(\ref{eq:current_relation}\right)$ have been
widely used by physicists and engineers for describing capacitors with time-varying capacitances. Unfortunately, this has been overlooked in the time-varying circuit theory~\cite{Richards2012}, as well as in the basic
circuit theory~\cite{Desoer2010}. Furthermore, the simulation tools such as those from Matlab and Micro-Cap~\cite{Biolek2007} also use Eq.~\ref{eq:current_relation} to model current through time-varying capacitors, which seems questionable. 

\subsection{Modification of classical-charge voltage relation}\label{sec:Revb}

The time-varying nature of the capacitance was already known to the industrial manufacturers of capacitors, but in lack of a better model than that expressed by Eq.~\ref{eq:charge_relation}, they defined capacitance at $1$ kHz~\cite{Westerlund1991}. A meticulous attempt was made by Westerlund to resolve this problem by proposing a charge--voltage relation for a universal capacitor
model~\cite{Westerlund1991}, but contrary to the expectation that work did not attract significant interest from the electrical engineers community because of three reasons. First, the lack of a closed-form expression for
the charge--voltage relation, see the abstract and Eqs.~(16) and (17) in Ref.~\cite{Westerlund1991}. Second, although his results were motivated from the experimental observations, the underlying issue with the classical
charge--voltage relation was not discussed. It is to be noted that this has now been addressed in the previous subsection. Third, the acceptability of the fractional derivatives among the scientific community was relatively
less three decades ago than it is now~\cite{Podlubny2002,Machado2015}. 

We propose the following equation as the charge--voltage relation of time-varying capacitors:
\begin{equation}
Q\left(t\right)=C\left(t\right)\ast\dot{V}\left(t\right),\label{eq:new_charge}
\end{equation}
 where $*$ represents the convolution operation.  Since the system is time-varying, yet linear, the derivative property of the convolution is applicable. Substituting  Eq.~\ref{eq:time_var_cap} in  Eq.~\ref{eq:new_charge} and following the derivative property of the convolution, we obtain the capacitor current as:
\begin{equation}
I\left(t\right)=C\left(t\right)\ast\ddot{V}\left(t\right)=\left[C_{0}\ast\ddot{V}\left(t\right)\right]+\left[\phi t\ast\ddot{V}\left(t\right)\right].\label{eq:convolv_current}
\end{equation}
 It is worth noticing that the expressions for, $Q$ and  $I$, in Eqs.~$\left(\ref{eq:new_charge}\right)$ and $\left(\ref{eq:convolv_current}\right)$, are actually motivated from the fractional derivatives as they provide an appropriate mathematical framework to model memory-laden systems. The expression for the current, Eq.~\ref{eq:convolv_current},  when seen in light of Eq.~\ref{eq:frac_derivat}, gives,
\begin{equation}
I\left(t\right)=C_{0}\dot{V}\left(t\right)+\phi V\left(t\right).\label{eq:new_current}
\end{equation}
Here, an important observation is that a parallel combination of a constant capacitance capacitor, $C_0$, and a time-varying  capacitor, $C_{\phi}$, is equivalent to the parallel combination of the constant capacitance capacitor and a resistor whose resistance is given by $1/\phi$. Also assuming, $V=0$ at $t=0$, at any instant of time, $t>0$, $V\left(t\right)\equiv t \dot{V}\left(t\right)$, which when substituted back in Eq.~\ref{eq:new_current}, leads to,
\begin{equation}
\begin{split}
        I\left(t\right)=I_{C_{0}}\left(t\right)+I_{C_{\phi}}\left(t\right),\\
        \text{ where }  I_{C_{0}}\left(t\right) = C_{0}\dot{V}\left(t\right), \text{ and }  I_{C_{\phi}} = C_{\phi} \dot{V}\left(t\right),
\end{split}
    \label{eq:super_new_current}
\end{equation}
are the currents through the capacitors of capacitances, $C_{0}$ and $C_{\phi}$, respectively. It is also possible to obtain  Eq.~\ref{eq:super_new_current} from Eq.~\ref{eq:convolv_current}
using the standard convolution integral, but if the time-varying capacitance is expressed in the form of a power-law, then the fractional framework turns out to be a readily available tool for their analysis.  The
correctness of Eq.~\ref{eq:super_new_current} that stems from the modified charge--voltage relation can be confirmed in two independent ways. First, the application of Kirchhoff's current law to the parallel combination
of the capacitors in  Fig.~\ref{Time_capacitor} (b) yields the same expression as Eq.~\ref{eq:super_new_current}, which would also be the case if  the term,   $V\phi$, in Eq.~\ref{eq:curie_current}, is ignored. Since
the term, $V\phi$, has its origin in the last term, $V\left(t\right)\dot{C}\left(t\right)$, in  Eq.~\ref{eq:current_relation}, it may be concluded that both the terms should not be present in their respective equations. This has been
experimentally verified as well~\cite{Jadli2020}. Second, in contrast to Eq.~\ref{eq:curie_current},     Eq.~\ref{eq:super_new_current} does not contain the undesired ohmic current term and is therefore equivalent to
the current flowing in the circuit shown in  Fig.~\ref{Time_capacitor} (b). Thus the non-equivalence that arose due to the application of the conventional charge--voltage equation, Eq.~\ref{eq:charge_relation}, is resolved
through the proposed convolution equation, Eq.~\ref{eq:new_charge}.

Although we have assumed a linear time-varying capacitance, the proof that we have presented here can be generalized to all power-law forms of the time-varying capacitance using fractional derivatives. On replacing, $C\left(t\right)$, in  Eq.~\ref{eq:convolv_current} by $C_{0}\left(\tau/t\right)^{\alpha-1}/\Gamma\left(2-\alpha\right)$, and then interpreting the resulting equation in light of Eq.~\ref{eq:frac_derivat}, we obtain the expression for the current through a fractional capacitor as:
\begin{equation}
I_{C_{f}}\left(t\right)=C_{0}\tau^{\alpha-1}\left[\frac{t^{1-\alpha}}{\Gamma\left(2-\alpha\right)}\ast\ddot{V}\left(t\right)\right]=C_{f}\frac{d^{\alpha}}{dt^{\alpha}}V\left(t\right),\label{eq:last_frac_current}
\end{equation}
where, $C_{f}=C_{0}\tau^{\alpha-1}$. If the capacitance is assumed to be a constant, i.e.,~$C_{\phi}=0$, then results from Eqs.~$\left(\ref{eq:new_charge}\right)$ and $\left(\ref{eq:convolv_current}\right)$,,  reduce to the classical relations, $Q\left(t\right)=C_{0}V\left(t\right)$, and $I\left(t\right)=C_{0}\dot{V}\left(t\right)$,
respectively, which is expected for a time-invariant system. This can be witnessed from Eq.~\ref{eq:super_new_current} too. The same also mirrors from Eq.~\ref{eq:last_frac_current}, for $\alpha=1$, which
corresponds to a memoryless system. Since Eq.~\ref{eq:last_frac_current} is obtained from Eq.~\ref{eq:convolv_current} which in turn has its origin in Eq.~\ref{eq:new_charge}, it is asserted that
Eq.~\ref{eq:new_charge} corresponds to the charge--voltage relation of a fractional capacitor.  Interestingly, the modified charge--voltage relation can also be seen in a scattered, non closed-form, in Eqs.~$\left(16\right)$ and $\left(17\right)$ of Ref.~\cite{Westerlund1991}. In contrast to the dimensionally consistent modified charge--voltage relation introduced in this manuscript, other competing expressions lack clarity in dimensional consistency, for example, Eq.~(5)
in Ref.~\cite{Fouda2020}.  The implications of the proposed convolution relations can be guessed from the fact that they are inherently required in the modeling of dielectric
media~\cite{Jonscher2001,Luo2004,XU2004,Jameson2006,Ning2008}, supercapacitors~\cite{Allagui2018a,Allagui2018b}, and electrochemical capacitors~\cite{Martynyuk2018}. Hence, we regard this as an intermediate finding
presented in this manuscript.

\section{Curie-von Schweidler law from time-varying capacitance\label{sec:Deriv_CvS_law}}

\begin{figure}[h]
\begin{centering}
\includegraphics[width=0.5\columnwidth]{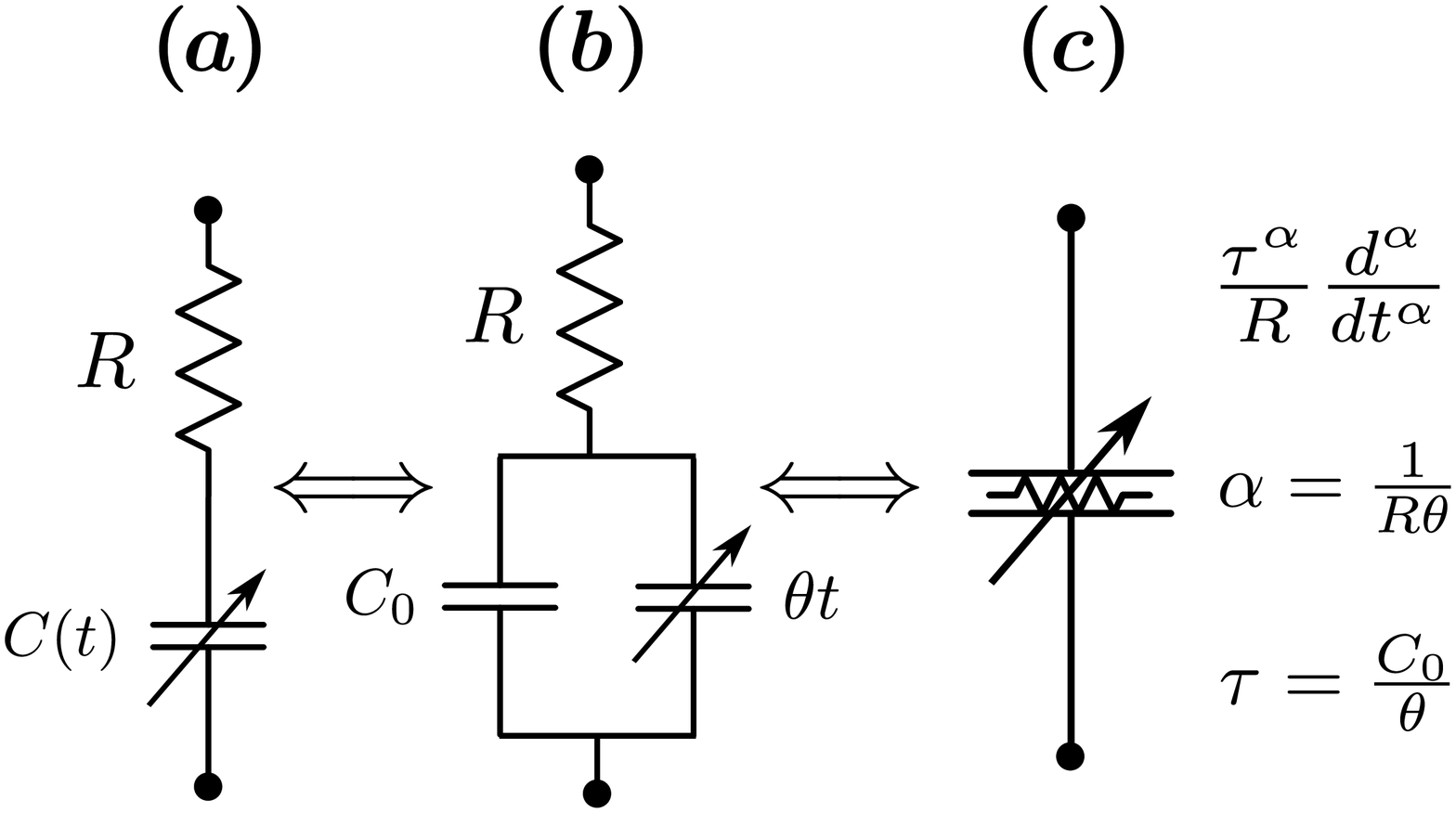}
\par\end{centering}
\caption{Equivalent circuit model of a fractional (\textit{universal}) capacitor. (a) Circuit consisting of a series combination of a resistor, $R$, and a time-dependent capacitor, $C\left(t\right)$. (b) Equivalent circuit of (a). (c) The proposed new symbol of the fractional capacitor whose current response is approximately the same as the current response of (b).}
\begin{centering}
\label{frac_capacitor}
\par\end{centering}
\end{figure}

Since the expression of the Nutting law~\cite{Pandey2016a} is quite similar to the expression of the CvS law, it is quite natural to follow the same route to derive the CvS law. Therefore, we replace the rheological modeling elements by their
respective electrical analogues~\cite{Giusti2016}, i.e., we replace the spring, the constant
viscosity dashpot, and the time-varying dashpot from the modified
Maxwell model of Fig.~1 (a) from Ref.~\cite{Pandey2016a}, by a resistor
of resistance $R$, a capacitor of constant capacitance, $C_{0}$,
and a capacitor of time-varying capacitance, $C_{\chi}\left(t\right)$,
respectively. Since
individual capacitances add in parallel circuits, the net capacitance,
$C\left(t\right)$, becomes
\begin{equation}
C\left(t\right)=C_{0}+C_{\chi}\left(t\right),\text{ }C_{\chi}\left(t\right)=\theta t,\text{ and }\theta=\left.\frac{dC\left(t\right)}{dt}\right|_{t=0}>0,\label{eq:capaci}
\end{equation}
where $C_{0}$ is the constant geometric capacitance, $C_{\chi}\left(t\right)$ is the linear time-varying capacitance due to the dielectric
material of the capacitor, and $\theta$ is a positive constant. The equivalent circuit model is shown in  n Figs.~\ref{frac_capacitor} (a) and (b), in which $R$ and $C\left(t\right)$, are considered to have their origin from the bulk dielectric and the time-dependent 
barrier respectively~\cite{Jonscher1983}. Since in actual experiments the charging and the discharging of a capacitor occurs through an external resistor connected in series with it, the resistance $R$ actually
represents the series combination of the resistance of the bulk dielectric and the external resistance. Although lead and plate
resistances are also present, but since they are often of the order of a few milli-ohms, they can be safely ignored in comparison to the external resistor whose resistance is usually of the order of a few kilo-ohms or even greater. The assumption of
a time-varying capacitance is not \textit{ad hoc}, it manifests from
a delayed-response that arises due to a time-dependent charge carrier
distribution in dielectrics~\cite{Schubert1985,Nowick1998,Kytin2001}.  Such an assumption has also been used
to harness energy~\cite{Mirmoosa2019} and to describe properties of
cerebral microvasculature~\cite{Jo2015}. 

A unit step voltage, $V_{0}$, is applied to the circuit in  Fig.~\ref{frac_capacitor} (b), at, $t=0$,
such that $I_{R}$, $I_{C}$, $V_{R}$, and $V_{C}$, are the currents
and the voltages across the resistor and the capacitor respectively. The voltage, $V_{C}\left(t\right)$, is the same in the parallel branches that have the two capacitors. Since the time-varying capacitance yields a time-varying
system, the modified charge--voltage equation, Eq.~\ref{eq:new_charge}, is used. Therefore, the current through the parallel combinations of the capacitors is, $I_{C}=I_{C_{0}}+I_{C_{\chi}}=C_{0}\dot{V}_{C}+\theta t\dot{V}_{C}$, which yields the respective voltage as,
$\dot{V}_{C}=I_{C}/\left(C_{0}+\theta t\right)$. This expression can also be directly extracted from Eq.~$\left(17\right)$ of Ref.~\cite{Westerlund1991}. In a series arrangement, $I_{R}=I_{C}=I$, and $V_{0}=V_{R}+V_{C}$.
Since, $\dot{V}_{0}=\dot{V}_{R}+\dot{V}_{C}=0$ and $\dot{V}_{R}=\dot{I}_{R}R$, we have,
\begin{equation}
R\dot{I}+\frac{I}{C_{0}+\theta t}=0.\label{eq:diffequat}
\end{equation}
The integration of Eq.~\ref{eq:diffequat} yields,
\begin{equation}
R\ln I=-\frac{\ln\left(C_{0}+\theta t\right)}\theta+\ln K.\label{eq:integresult}
\end{equation}
Further, at $t=0$, the current flows only through the resistance,
i.e.,~$I\left(t=0\right)=I_{0}=V_{0}/R$. On imposing this initial condition,
the integration constant, $\ln K$, is obtained as, $\ln K=R\ln I_{0}+\left(\ln C_{0}\right)/\theta$,
which on substituting back into the Eq.~\ref{eq:integresult}, yields,
\begin{equation}
I\left(t\right)=I_{0}\left(1+\frac{\theta}{C_{0}}t\right)^{-1/\left(R\theta\right)}.\label{eq:Result_exp1}
\end{equation}
If the time-varying part of the capacitance dominates over the constant capacitance part, i.e.,~if $C_{\chi}\left(t\right) = \theta t \gg C_{0}$, then Eq.~\ref{eq:Result_exp1}
reduces to,
\begin{equation}
I\left(t\right)\approx I_{0}\left(\frac{\theta}{C_{0}}t\right)^{-G/\theta},\label{eq:Result_exp2}
\end{equation}
which is the CvS law. On comparing Eq.~\ref{eq:Result_exp2}
with Eqs.~$\left(\ref{eq:cvslaw}\right)$ and $\left(\ref{eq:frac_cap}\right)$, the parameters of the UDR
and the fractional capacitor gain physical interpretation as:
\begin{equation}
a=\frac{V_{0}}{R},\text{ }\text{ }\alpha=\frac{G}{\theta},\text{ }\tau=\frac{C_{0}}{\theta},\text{ }C_{f}=G\left(\frac{C_{0}}{\theta}\right)^{\alpha}, \text{ where } G=\frac{1}{R},\label{eq:Result_parameter}
\end{equation}
may be seen as the total electrical conductance present in the circuit. Further since the fractional capacitor satisfies the Kramers--Kronig relations~\cite{Westerlund1994}, this naturally extends to the circuit model shown
in  Fig.~\ref{frac_capacitor} (a). A new
symbol for the fractional capacitor is proposed in  Fig.~\ref{frac_capacitor} (c)  to
distinguish it from the symbol used in  Figs.~\ref{frac_capacitor} (a) and (b) that has traditionally been used to represent the varying-capacitance due to the voltage-induced nonlinearity. The new
symbol contains the symbol of a resistor sandwiched in between the parallel-plate symbol of a capacitor.

It can be seen from Eqs.~$\left(\ref{eq:Result_exp2}\right)$ and $\left(\ref{eq:Result_parameter}\right)$  that as $\alpha\rightarrow0$, $I\rightarrow I_{0}$, and $C_{f}\rightarrow 1/R$. This means the dielectric behaves purely resistive
in accordance with the Ohm law which can also be verified from Eq.~\ref{eq:frac_cap}. However, as $\alpha\rightarrow1$, there are two possible outcomes that are dependent on the magnitude of the ratio, $t/\tau$. On the one hand, as $t/\tau\rightarrow\infty$, it implies, $\tau\rightarrow 0$, and so, $\theta\rightarrow \infty$, which in turn yields, $C\left(t\right)\rightarrow\infty$. This is in agreement with the observation for polypropylene for which
$1>\alpha>0.999$~\cite{Westerlund1991}. At such long timescales,  $I\rightarrow 0$, and this can be verified from Eq.~\ref{eq:Result_exp2}  as well as from Eq.~\ref{eq:frac_cap}, where  $C_{f}\rightarrow 0$. On the other hand, we can use Euler's limit definition of the exponential function, in Eq.~\ref{eq:Result_exp1},
to give an exponential decaying current, $I\rightarrow I_{0}\exp\left(-t/\tau\right)$, at short timescales, i.e.,~as $\tau/t\rightarrow\infty$. This is expected from an ideal memoryless capacitor because such a short timescale necessitates, $\theta\rightarrow0$. Alternatively, there is no significant contribution to the capacitance from its time-varying part, $C_{\chi}\left(t\right)$, at short timescales. This implies that the circuit model in Fig.~\ref{frac_capacitor} (b) is then reduced to the series combination of $R$ and $C_{0}$, which is actually the equivalent circuit for the Debye response. This is also verifiable, since as, $\alpha=1/\left(R\theta\right)\rightarrow1$,
$\theta\rightarrow1/R$, which leads to, $\tau=C_{0}/\theta\rightarrow RC_{0}=\tau_{D}$. Therefore, it is inferred that a non-Debye dielectric medium appears as a Debye medium at very short
timescales. To summarize, the case, $\alpha\rightarrow1$, hints at the presence of
two different relaxations time constants, the Debye relaxation time constant, $\tau_{D}=RC_{0}$, at short
timescales, and the non-Debye relaxation time constant, $\tau=C_{0}/\theta$, at long timescales. It is
therefore concluded that the Debye relaxation mechanism and the non-Debye
relaxation mechanism are two independent mechanisms
that coexist and occur sequentially in a dielectric. This is in accordance with the experimental observations~\cite{Jonscher1983,Bunde1997,Novikov2001,Raicu2001}, which  may also possibly explain the sensitivity of the
curve-fitting parameters to the
frequency window employed in the analysis~\cite{Ngai1996}. The transition from the Debye relaxation to the non-Debye relaxation is termed as the ``\textit{regenerated memory effect}'', and has already been experimentally
observed, see Figs.~[4] and [5] in Ref.~\cite{Uchaikin2009}. In the frequency domain, the high-frequency dispersion
due to the Debye response can be seen as a natural extension of the low frequency
dispersion due to the UDR. This is in agreement with the published reports~\cite{Bunde1997,Raicu1999,Novikov2001,Uchaikin2009}. Since the revelation of the hidden Debye response comes from Eq.~\ref{eq:Result_exp1}, it should be rather preferred as the expression for the UDR instead of its approximated version, Eq.~\ref{eq:Result_exp2}.

\begin{table}[H]
\centering
\begin{tabular}{|c|c|c|c|c|@{}p{0cm}|c|c|}
\hline 
\multicolumn{5}{|c|}{Experimental data} &  & \multicolumn{2}{c|}{Theoretical predictions from Eq.$\left(\ref{eq:Result_parameter}\right)$}\tabularnewline
\hline 
Source & $C_{0}$ & $R$ & $\tau$ & $\alpha$  &  & $\theta=C_{0}/\tau$ & $\alpha=1/\left(R\theta\right)$\tabularnewline
\hline 
Ref.~\cite[Fig.~5]{Westerlund1994} & $1\text{ \ensuremath{\mu}F}$ & $1\text{ M\ensuremath{\Omega}}$ & $1\text{ $s$}$ & $0.982$ & & $1\text{ \ensuremath{\mu}F}/s$ & $1$ \tabularnewline
\hline 
Ref.~\cite[Fig.~4]{Uchaikin2009} & $2\text{ \ensuremath{\mu}F}$ & $200\text{ k\ensuremath{\Omega}}$ & $0.4\text{ $s$}$ & $0.998$ & & $5.01\text{ \ensuremath{\mu}F}/s$ & $0.998$ \tabularnewline
\hline 
\end{tabular}
\caption{Comparison of the experimental data against the theoretical predictions from Eq.$\left(\ref{eq:Result_parameter}\right)$.}
\label{capacitor_data}
\end{table}

In order to further consolidate our findings, we compare them with the experimental data in 
Table~\ref{capacitor_data}. The charging of a capacitor consisting of metalized paper was investigated in  Ref.~\cite{Westerlund1994}, in contrast, the charging of a capacitor consisting of oil-impregnated technical paper
was studied in  Ref.~\cite{Uchaikin2009}. Since the experimental data which is available for comparison is of capacitors with values of $\alpha \approx 1$, it implies that those capacitors exhibit almost an ideal capacitive behavior. In such cases, the resistance due to the bulk dielectric is expected to be negligible. Consequently, the resistance, $R$, effectively represents the resistance of the external resistor  that
is connected in series with the capacitor. A reasonably good agreement between the experimental observations and the predictions from Eq.~\ref{eq:Result_parameter} can be witnessed for the value of the power-law
exponent, $\alpha$. Moreover, the predicted value of the current, $I_{0}=V_{0}/R=10^{-3}\text{ A}$, is almost identical to the observation in which the applied voltage bias was, $V_{0}=200 \text { V}$, see  Ref.~\cite[Fig.~4]{Uchaikin2009}.

\section{Discussion\label{sec:Discussion}}

We have achieved three goals here. \textit{First}, the time-honored
UDR is identified as an electrical analogue of Nutting's law. This asserts the effectiveness of fractional derivatives in modeling of materials that exhibit memory. The
derivation and the subsequent analysis presented here is in accordance
with Jonscher's prediction of a satisfactory theory that is independent
of the detailed physical and chemical nature of the dielectrics. The assumption of a time-varying capacitance underlying the derivation of the CvS law is possibly favored by the principle of parsimony too. \textit{Second}, the parameters of the UDR have gained physical interpretation and therefore it may reduce ambiguities in curve-fits. Since the fractional
capacitor has its origin in the UDR, the benefits automatically extend
to those circuit models that have fractional derivatives in them~\cite{Gomez2016}. \textit{Third}, it is shown that a time-varying capacitor is governed by the equation, $Q\left(t\right)=C\left(t\right)\ast\dot{V}\left(t\right)$. The consequent predictions match quite well with the experimental results. Interestingly, the convolution relation reduces to the classical relation, $Q\left(t\right)=CV\left(t\right)$, if the capacitance is assumed to be a constant. Therefore, we believe the former should be preferred over the latter as that is applicable for both memoryless as well as for memory-laden dielectric media. The convolution equations should be seen as relations that complete the big picture and yet retain the beauty of the classical relations. Moreover, since the fractional capacitor is also considered as the universal capacitor, the relation, $Q\left(t\right)=C\left(t\right)\ast\dot{V}\left(t\right)$, may be regarded as the charge--voltage relation of a \textit{universal capacitor}. It is expected that this finding may further boost the emerging field of fractional-order circuits and systems. 

Although phenomenological models have been used here, it is shown that a careful
analysis of them could facilitate a  cross-fertilization of the disparate
fields of non-Newtonian rheology and dielectric spectroscopy in the
framework of fractional calculus.  The preference for the fractional derivatives in describing  complex media that exhibit memory is similar in principle to the preference given to an appropriate
coordinate system that suits the geometry of the physical phenomena under investigation. We emphasize that fractional calculus is not just
a mathematical tool that can only be used to curve-fit the anomalous behavior
of complex media. Rather, it has an inherent connection with physics
too that needs to be explored more. 

Lastly, we accept a possible critique of this work, which is that the model presented here is macroscopic and is probably not reducible to a microscopic level. A similar yet a contrasting notion is that if a macroscopic
model can satisfactorily describe an experimentally established physical law, then probably the difference between a macroscopic description and a microscopic description is merely an ideological one.  We would like to stress upon the
idea that the arising of memory in a material may be regarded as an example of \textit{weak emergence}. This is actually in accordance with the finding that even if a perfect, complete description of the microscopic interactions is available, it may not always be sufficient to deduce all of the macroscopic properties of a material~\cite{Cubitt2015}. The idea of emergence should not be seen here as an excuse for the
lack of knowledge, rather we are optimistic and confident about the explanatory powers of science.

\section*{Acknowledgment\label{sec:Acknowledgement}}
The author would like to thank Prof.~Sivakumar Srinivasan of Krea University, Sricity, for going through the manuscript
and for the fruitful discussions that the author had with him.



\end{document}